\documentstyle[aps,graphicx,floats,preprint]{revtex}

\begin{document}

\title{Spectroscopic evidence for unconventional superconductivity in 
UBe$_{13}$}

\author{Ch. W\"alti and H. R. Ott}
\address{Laboratorium f\"ur Festk\"orperphysik, ETH-H\"onggerberg,
          8093 Z\"urich, Switzerland}
\author{Z. Fisk}
\address{National High Magnetic Field Laboratory, Tallahassee, FL 32306-4005,
USA }
\author{J. L. Smith}
\address{Los Alamos National Laboratory, Los Alamos,
New Mexico 87545, USA}

\maketitle

In a recent letter~\cite{Letter}, we reported the observation of 
giant zero-bias conductance peaks (ZBCPs) for contacts between Au 
wires and bulk UBe$_{13}$, which we ascribed to the 
existence of low-energy Andreev surface bound-states in  
superconducting UBe$_{13}$.
In the comment~\cite{Gloos} preceding this reply,
Gloos proposes an alternative interpretation of the ZBCPs observed in 
Ref.~\onlinecite{Letter} which is based on 
Wexler's formula~\cite{Gloos}.
Below, we  first show
that the criticism of our interpretation by Gloos is not relevant for the
experimental situation discussed in Ref.~\onlinecite{Letter} and second, we 
argue that the
observed ZBCPs in the UBe$_{13}$--Au contacts are not compatible with the
interpretation given in Ref.~\onlinecite{Gloos}.

In Ref.~\onlinecite{Gloos}, Gloos discusses the properties of 
normal metal--normal metal (NN) contacts by means of  Wexler's 
formula, from which
he derives a condition claiming that the UBe$_{13}$--Au 
contacts discussed in Ref.~\onlinecite{Letter} are not in the limit  where pure 
Andreev reflection (AR)~\cite{2498} may occur. 
Pure AR, or electron-hole reflection, at the interface between a 
normal conductor and a superconductor (NS) 
may cause an enhancement of the differential conductivity 
$G(E)$ for $|E|<\Delta$, where $\Delta$ denotes the amplitude of the superconducting 
energy gap, but 
only in the limit of a low potential-barrier at the interface~\cite{2487}. The 
mean free path $l$ in the contact, which is  discussed in Ref.~\onlinecite{Gloos},
is large, if the potential-barrier is small and vice versa. 
However, in our discussion of the observed ZBCPs we do not claim to observe an 
enhancement of the differential conductivity of the contact due to pure AR.
Instead we claim 
to observe a surface resonance phenomenon, which is caused by the presence of 
subgap Andreev bound-states. These 
two AR-type phenomena are simply not the same.
 
Wexler's formula, in the form used in Ref.~\onlinecite{Gloos}, is,
apart from other criteria, based on
the assumption that the Fermi-liquid parameters and hence the 
resistivity $\rho$ of both metals on
either side of the contact are at least similar. 
Since in Ref.~\onlinecite{Letter} one of the metals of the
contact is the heavy-electron metal UBe$_{13}$, this assumption is
certainly not valid, and Wexler's formula in the form as employed by
Gloos~\cite{Gloos} needs to be revised. 
In addition, the application of Wexler's formula for estimating 
the contact radius $a$ using the contact 
resistance $R$ and $\rho$ of  bulk UBe$_{13}$ is, in our opinion, an 
invalid step.

The interpretation of our experiments~\cite{Letter} which is 
proposed in Ref.~\onlinecite{Gloos} is based on the
assumption that the loss of the electrical resistance of UBe$_{13}$ in its 
superconducting state is responsible for the huge enhancement of the differential 
conductivity at zero energy. 
The discussion given in  Ref.~\onlinecite{Gloos} only embraces the temperature 
dependence of the conductivity of NN contacts. Discussing the 
{\em energy}-dependent differential conductivity $G(E)$ of, e.g., UBe$_{13}$--Au 
contacts, however, would require a model including its  {\em energy} dependence.
This important issue is not contained in the discussion presented in Ref.~\onlinecite{Gloos}.

It is well known that superconductivity of UBe$_{13}$ with zero 
electrical resistance is fully established 
within a few ten mK below $T_c$.
Therefore, in the context of Gloos' model, one would expect, for $|E|<\Delta$, the 
differential conductivity of the UBe$_{13}$--Au contact 
to increase  sharply at $T_{c}$, reaching
its maximum just below $T_c$. 
In that case there would be no obvious reason why 
the differential conductivity at zero energy, $G(0)$, should further increase
substantially with decreasing temperature, as it is observed for 
the data discussed in Ref.~\onlinecite{Letter}.

The shape of the $G(E)$ curve of the UBe$_{13}$--Au contact (see Fig.~3 of 
Ref.~\onlinecite{Letter}) does also not match the predictions of the model proposed by 
Gloos. The electrical resistance of UBe$_{13}$ at $T<T_c$ is zero for 
$|E|<\Delta$. The model discussed in Ref.~\onlinecite{Gloos} does not offer an 
explanation, why $G(E)$ of the UBe$_{13}$--Au contact is only strongly 
enhanced in a narrow region around zero bias, but substantially 
{\em reduced} compared to the normal state of the contact at intermediate 
energies $|E|<\Delta$.

We acknowledge helpful discussions with M. Sigrist.

\end{document}